

\input epsf


\def\prd#1#2#3#4{#1, {\it Phys. Rev. \/}{\bf D#2} (19#4) #3}
\def\jetf#1#2#3#4{#1, {\it JETP Letters \/}{\bf #2} (19#4) #3}
\def\pt#1#2#3#4{#1, {\it Physics Today \/}{\bf #2} (19#4) #3}
\def\pr#1#2#3#4{#1, {\it Phys. Rev. \/}{\bf #2} (19#4) #3}
\def\cmp#1#2#3#4{#1, {\it Commun. Math. Phys. \/}{\bf #2} (19#4) #3}
\def\prl#1#2#3#4{#1, {\it Phys. Rev. Lett.\/ }{\bf #2} (19#4) #3}
\def\pla#1#2#3#4{#1, {\it Phys. Lett. \/}{\bf A#2} (19#4) #3}

\def\npb#1#2#3#4{#1, {\it Nucl. Phys. \/}{\bf B#2} (19#4) #3}
\def\ijmpc#1#2#3#4{#1, {\it Int. J. Mod. Phys. \/}{\bf C#2} (19#4) #3}

\def\ncl#1#2#3#4{#1, {\it Lett. Nuov. Cimento \/}{\bf #2} (19#4) #3}

\def\nat#1#2#3#4{#1, {\it Nature (Physical Science) \/}{\bf #2} (19#4) #3}
\def\natps#1#2#3#4{#1, {\it Nature \/}{\bf #2} (19#4) #3}
\def\preprint#1#2#3#4{#1, ``#2'', #3 (19#4)}

\def\inbooked#1#2#3#4#5{#1, in {\it #2\/,} ed. #3 ({#4,} 19#5)}
\def\inbookeds#1#2#3#4#5{#1, in {\it #2\/,} eds. #3  ({#4,} 19#5)}
\def\book#1#2#3#4{#1,{\it #2\/} (#3, 19#4)}

\font\twelverm=cmr10 scaled 1200    \font\twelvei=cmmi10 scaled 1200
\font\twelvesy=cmsy10 scaled 1200   \font\twelveex=cmex10 scaled 1200
\font\twelvebf=cmbx10 scaled 1200   \font\twelvesl=cmsl10 scaled 1200
\font\twelvett=cmtt10 scaled 1200   \font\twelveit=cmti10 scaled 1200

\skewchar\twelvei='177   \skewchar\twelvesy='60


\def\twelvepoint{\normalbaselineskip=12.4pt
  \abovedisplayskip 12.4pt plus 3pt minus 9pt
  \belowdisplayskip 12.4pt plus 3pt minus 9pt
  \abovedisplayshortskip 0pt plus 3pt
  \belowdisplayshortskip 7.2pt plus 3pt minus 4pt
  \smallskipamount=3.6pt plus1.2pt minus1.2pt
  \medskipamount=7.2pt plus2.4pt minus2.4pt
  \bigskipamount=14.4pt plus4.8pt minus4.8pt
  \def\rm{\fam0\twelverm}          \def\it{\fam\itfam\twelveit}%
  \def\sl{\fam\slfam\twelvesl}     \def\bf{\fam\bffam\twelvebf}%
  \def\mit{\fam 1}                 \def\cal{\fam 2}%
  \def\tt{\twelvett}
  \def\nullspace{\nulldelimiterspace=0pt \mathsurround=0pt }
  \def\big##1{{\hbox{$\left##1\vbox to 10.2pt{}\right.\nullspace$}}}
  \def\Big##1{{\hbox{$\left##1\vbox to 13.8pt{}\right.\nullspace$}}}
  \def\bigg##1{{\hbox{$\left##1\vbox to 17.4pt{}\right.\nullspace$}}}
  \def\Bigg##1{{\hbox{$\left##1\vbox to 21.0pt{}\right.\nullspace$}}}
  \textfont0=\twelverm   \scriptfont0=\tenrm   \scriptscriptfont0=\sevenrm
  \textfont1=\twelvei    \scriptfont1=\teni    \scriptscriptfont1=\seveni
  \textfont2=\twelvesy   \scriptfont2=\tensy   \scriptscriptfont2=\sevensy
  \textfont3=\twelveex   \scriptfont3=\twelveex  \scriptscriptfont3=\twelveex
  \textfont\itfam=\twelveit
  \textfont\slfam=\twelvesl
  \textfont\bffam=\twelvebf
  \scriptfont\bffam=\tenbf
  \scriptscriptfont\bffam=\sevenbf
  \normalbaselines\rm}



\def\beginlinemode{\endmode
  \begingroup\parskip=0pt \obeylines\def\\{\par}\def\endmode{\par\endgroup}}
\def\beginparmode{\endmode
  \begingroup \def\endmode{\par\endgroup}}
\let\endmode=\par
{\obeylines\gdef\
{}}
\def\singlespace{\baselineskip=\normalbaselineskip}
\def\oneandahalfspace{\baselineskip=\normalbaselineskip
  \multiply\baselineskip by 3 \divide\baselineskip by 2}

\def\doublespace{\baselineskip=\normalbaselineskip \multiply\baselineskip by 2}
\newcount\firstpageno
\firstpageno=2
\footline={\ifnum\pageno<\firstpageno{\hfil}\else{\hfil\twelverm\folio\hfil}\fi}
\let\rawfootnote=\footnote
\def\footnote#1#2{{\tenrm\singlespace\parindent=0pt\rawfootnote{#1}{#2}}}
\def\raggedcenter{\leftskip=4em plus 12em \rightskip=\leftskip
  \parindent=0pt \parfillskip=0pt \spaceskip=.3333em \xspaceskip=.5em
  \pretolerance=9999 \tolerance=9999
  \hyphenpenalty=9999 \exhyphenpenalty=9999 }
\def\dateline{\rightline{\ifcase\month\or
  January\or February\or March\or April\or May\or June\or
  July\or August\or September\or October\or November\or December\fi
  \space\number\year}}
\def\received{\vskip 3pt plus 0.2fill
 \centerline{\sl (Received\space\ifcase\month\or
  January\or February\or March\or April\or May\or June\or
  July\or August\or September\or October\or November\or December\fi
  \qquad, \number\year)}}


\hsize=16truecm 
\hoffset=0.0truecm
\vsize=25truecm
\voffset=0.0truecm
\parskip=0pt
\twelvepoint  
\overfullrule=0pt 



\def\title
  {\null\vskip 3pt plus 0.2fill
   \beginlinemode \doublespace \raggedcenter \bf}


\font\twelvesc=cmcsc10 scaled 1200
\def\author{\vskip 16pt plus 0.2fill \beginlinemode\singlespace
\raggedcenter\twelvesc}

\def\affil
  {\vskip 4pt plus 0.1fill \beginlinemode
   \oneandahalfspace \raggedcenter \sl}

\def\abstract  
  {\vskip 24pt plus 0.3fill \beginparmode
   \narrower\centerline{ABSTRACT}\vskip 12pt }

\def\endpage{\vfill\eject}     

\def\body{\beginparmode}


\def\head#1{
  \filbreak\vskip 0.35truein
  {\immediate\write16{#1}
   \raggedcenter \uppercase{#1}\par}
   \nobreak\vskip 0.2truein\nobreak}

\def\References  
  {
   \beginparmode
   \frenchspacing\parindent=0pt \leftskip=0.5truecm
   \parskip=2pt plus 3pt \everypar{\hangindent=\parindent}}

\gdef\refis#1{\indent\hbox to 0pt{\hss#1.~}}.   

\def\endreferences{\body}


\def\figurecaptions
  {\endpage
   \beginparmode
   \head{Figure Captions}
}



\def\endpaper{\endmode\vfill\supereject}
\def\endit{\endpaper\end}


\def\cite#1{{#1}}
\def\[#1]{[\cite{#1}]}       
\def\refto#1{$^{{#1}}$}
\def\ref#1{Ref.~#1}          
\def\Ref#1{Ref.~#1}          

\def\call#1{{#1}}
\def\(#1){(\call{#1})}
\def\Eq#1{Eq.~\(#1)}                   

\catcode`@=11
\newcount\tagnumber\tagnumber=0

\immediate\newwrite\eqnfile
\newif\if@qnfile\@qnfilefalse
\def\write@qn#1{}
\def\writenew@qn#1{}
\def\w@rnwrite#1{\write@qn{#1}\message{#1}}
\def\@rrwrite#1{\write@qn{#1}\errmessage{#1}}

\def\t@ghead{}
\def\taghead#1{\gdef\t@ghead{#1}\global\tagnumber=0}

\expandafter\def\csname @qnnum-3\endcsname
  {{\t@ghead\advance\tagnumber by -3\relax\number\tagnumber}}
\expandafter\def\csname @qnnum-2\endcsname
  {{\t@ghead\advance\tagnumber by -2\relax\number\tagnumber}}
\expandafter\def\csname @qnnum-1\endcsname
  {{\t@ghead\advance\tagnumber by -1\relax\number\tagnumber}}
\expandafter\def\csname @qnnum0\endcsname
  {\t@ghead\number\tagnumber}
\expandafter\def\csname @qnnum+1\endcsname
  {{\t@ghead\advance\tagnumber by 1\relax\number\tagnumber}}
\expandafter\def\csname @qnnum+2\endcsname
  {{\t@ghead\advance\tagnumber by 2\relax\number\tagnumber}}
\expandafter\def\csname @qnnum+3\endcsname
  {{\t@ghead\advance\tagnumber by 3\relax\number\tagnumber}}

\def\equationfile{%
  \@qnfiletrue\immediate\openout\eqnfile=\jobname.eqn%
  \def\write@qn##1{\if@qnfile\immediate\write\eqnfile{##1}\fi}
  \def\writenew@qn##1{\if@qnfile\immediate\write\eqnfile
    {\noexpand\tag{##1} = (\t@ghead\number\tagnumber)}\fi}
}

\def\callall#1{\xdef#1##1{#1{\noexpand\call{##1}}}}
\def\call#1{\each@rg\callr@nge{#1}}

\def\each@rg#1#2{{\let\thecsname=#1\expandafter\first@rg#2,\end,}}
\def\first@rg#1,{\thecsname{#1}\apply@rg}
\def\apply@rg#1,{\ifx\end#1\let\next=\relax%
\else,\thecsname{#1}\let\next=\apply@rg\fi\next}

\def\callr@nge#1{\calldor@nge#1-\end-}
\def\callr@ngeat#1\end-{#1}
\def\calldor@nge#1-#2-{\ifx\end#2\@qneatspace#1 %
  \else\calll@@p{#1}{#2}\callr@ngeat\fi}
\def\calll@@p#1#2{\ifnum#1>#2{\@rrwrite{Equation range #1-#2\space is bad.}
\errhelp{If you call a series of equations by the notation M-N, then M and
N must be integers, and N must be greater than or equal to M.}}\else%
 {\count0=#1\count1=#2\advance\count1
by1\relax\expandafter\@qncall\the\count0,%
  \loop\advance\count0 by1\relax%
    \ifnum\count0<\count1,\expandafter\@qncall\the\count0,%
  \repeat}\fi}

\def\@qneatspace#1#2 {\@qncall#1#2,}
\def\@qncall#1,{\ifunc@lled{#1}{\def\next{#1}\ifx\next\empty\else
  \w@rnwrite{Equation number \noexpand\(>>#1<<) has not been defined yet.}
  >>#1<<\fi}\else\csname @qnnum#1\endcsname\fi}

\let\eqnono=\eqno
\def\eqno(#1){\tag#1}
\def\tag#1$${\eqnono(\displayt@g#1 )$$}

\def\aligntag#1\endaligntag
  $${\gdef\tag##1\\{&(##1 )\cr}\eqalignno{#1\\}$$
  \gdef\tag##1$${\eqnono(\displayt@g##1 )$$}}

\def\eqalignno#1{\displ@y \tabskip\centering
  \halign to\displaywidth{\hfil$\displaystyle{##}$\tabskip\z@skip
    &$\displaystyle{{}##}$\hfil\tabskip\centering
    &\llap{$\displayt@gpar##$}\tabskip\z@skip\crcr
    #1\crcr}}

\def\displayt@gpar(#1){(\displayt@g#1 )}

\def\displayt@g#1 {\rm\ifunc@lled{#1}\global\advance\tagnumber by1
        {\def\next{#1}\ifx\next\empty\else\expandafter
        \xdef\csname @qnnum#1\endcsname{\t@ghead\number\tagnumber}\fi}%
  \writenew@qn{#1}\t@ghead\number\tagnumber\else
        {\edef\next{\t@ghead\number\tagnumber}%
        \expandafter\ifx\csname @qnnum#1\endcsname\next\else
        \w@rnwrite{Equation \noexpand\tag{#1} is a duplicate number.}\fi}%
  \csname @qnnum#1\endcsname\fi}

\def\ifunc@lled#1{\expandafter\ifx\csname @qnnum#1\endcsname\relax}

\let\@qnend=\end\gdef\end{\if@qnfile
\immediate\write16{Equation numbers written on []\jobname.EQN.}\fi\@qnend}

\catcode`@=12


\catcode`@=11
\newcount\r@fcount \r@fcount=0
\newcount\r@fcurr
\immediate\newwrite\reffile
\newif\ifr@ffile\r@ffilefalse
\def\w@rnwrite#1{\ifr@ffile\immediate\write\reffile{#1}\fi\message{#1}}

\def\writer@f#1>>{}
\def\referencefile{
  \r@ffiletrue\immediate\openout\reffile=\jobname.ref%
  \def\writer@f##1>>{\ifr@ffile\immediate\write\reffile%
    {\noexpand\refis{##1} = \csname r@fnum##1\endcsname = %
     \expandafter\expandafter\expandafter\strip@t\expandafter%
     \meaning\csname r@ftext\csname r@fnum##1\endcsname\endcsname}\fi}%
  \def\strip@t##1>>{}}

\def\citeall#1{\xdef#1##1{#1{\noexpand\cite{##1}}}}
\def\cite#1{\each@rg\citer@nge{#1}}

\def\each@rg#1#2{{\let\thecsname=#1\expandafter\first@rg#2,\end,}}
\def\first@rg#1,{\thecsname{#1}\apply@rg}
\def\apply@rg#1,{\ifx\end#1\let\next=\relax
\else,\thecsname{#1}\let\next=\apply@rg\fi\next}

\def\citer@nge#1{\citedor@nge#1-\end-}
\def\citer@ngeat#1\end-{#1}
\def\citedor@nge#1-#2-{\ifx\end#2\r@featspace#1 
  \else\citel@@p{#1}{#2}\citer@ngeat\fi}
\def\citel@@p#1#2{\ifnum#1>#2{\errmessage{Reference range #1-#2\space is bad.}%
    \errhelp{If you cite a series of references by the notation M-N, then M and
    N must be integers, and N must be greater than or equal to M.}}\else%
 {\count0=#1\count1=#2\advance\count1
by1\relax\expandafter\r@fcite\the\count0,%
  \loop\advance\count0 by1\relax
    \ifnum\count0<\count1,\expandafter\r@fcite\the\count0,%
  \repeat}\fi}

\def\r@featspace#1#2 {\r@fcite#1#2,}
\def\r@fcite#1,{\ifuncit@d{#1}
    \newr@f{#1}%
    \expandafter\gdef\csname r@ftext\number\r@fcount\endcsname%
                     {\message{Reference #1 to be supplied.}%
                      \writer@f#1>>#1 to be supplied.\par}%
 \fi%
 \csname r@fnum#1\endcsname}
\def\ifuncit@d#1{\expandafter\ifx\csname r@fnum#1\endcsname\relax}%
\def\newr@f#1{\global\advance\r@fcount by1%
    \expandafter\xdef\csname r@fnum#1\endcsname{\number\r@fcount}}

\let\r@fis=\refis
\def\refis#1#2#3\par{\ifuncit@d{#1}
   \newr@f{#1}%
   \w@rnwrite{Reference #1=\number\r@fcount\space is not cited up to now.}\fi%
  \expandafter\gdef\csname r@ftext\csname r@fnum#1\endcsname\endcsname%
  {\writer@f#1>>#2#3\par}}

\def\ignoreuncited{
   \def\refis##1##2##3\par{\ifuncit@d{##1}%
     \else\expandafter\gdef\csname r@ftext\csname
r@fnum##1\endcsname\endcsname%
     {\writer@f##1>>##2##3\par}\fi}}

\def\r@ferr{\endreferences\errmessage{I was expecting to see
\noexpand\endreferences before now;  I have inserted it here.}}
\let\r@ferences=\references
\def\references{\r@ferences\def\endmode{\r@ferr\par\endgroup}}

\let\endr@ferences=\endreferences
\def\endreferences{\r@fcurr=0
  {\loop\ifnum\r@fcurr<\r@fcount
    \advance\r@fcurr by 1\relax\expandafter\r@fis\expandafter{\number\r@fcurr}%
    \csname r@ftext\number\r@fcurr\endcsname%
  \repeat}\gdef\r@ferr{}\endr@ferences}


\let\r@fend=\endpaper\gdef\endpaper{\ifr@ffile
\immediate\write16{Cross References written on []\jobname.REF.}\fi\r@fend}

\catcode`@=12

\def\reftorange#1#2#3{$^{\cite{#1}-\setbox0=\hbox{\cite{#2}}\cite{#3}}$}

\citeall\refto
\citeall\ref%
\citeall\Ref%

\ignoreuncited
\parindent=1.5truepc
\hsize=6.0truein
\vsize=9.6truein
\nopagenumbers

\line{\hfil gr-qc/9409015}
\medskip
\centerline{\tenbf DO WE UNDERSTAND BLACK HOLE ENTROPY ?
\footnote{$^*$}{Plenary
talk at Seventh Marcel Grossman Meeting, Stanford University, July 1994.  To
appear in the proceedings to be published by World Scientific.}}
\vglue 0.8cm
\centerline{\tenrm JACOB D. BEKENSTEIN}
\baselineskip=13pt
\centerline{\tenit Racah Institute of Physics, The Hebrew University of
Jerusalem}
\baselineskip=12pt
\centerline{\tenit Givat Ram, Jerusalem, 91904 ISRAEL}
\vglue 0.8cm
\centerline{\tenrm ABSTRACT}
\vglue 0.3cm
{\rightskip=3pc
 \leftskip=3pc
\tenrm\baselineskip=12pt\noindent
I review  various proposals for the nature of black hole entropy and for the
mechanism behind the operation of the generalized second law.  I stress the
merits  of entanglement entropy {\tenit qua\/} black hole entropy, and point
out that, from an operational viewpoint, entanglement entropy is perfectly
finite.  Problems with this identification such as the multispecies problem
and the trivialization of the information puzzle are mentioned.  This last
leads me to associate black hole entropy rather with the multiplicity of
density operators which describe a black hole according to exterior observers.
I  relate this identification to Sorkin's proof of the generalized second law.
I discuss in some depth Frolov and Page's proof of the same law,  finding it
relevant only for scattering of microsystems by a black hole.  Assuming that
the
law is generally valid I make evident the existence of the universal bound on
entropy regardless of issues of acceleration buoyancy, and discuss the
question of why macroscopic objects cannot emerge in the Hawking radiance.
\vglue 0.6cm}
\vfil
\twelverm

\baselineskip=14pt
\leftline{\twelvebf 1. Introduction}
\vglue 0.4cm

Three intricately related issues have characterized black hole thermodynamics
for the better part of two decades: the meaning of black hole entropy,  the
mechanism behind the operation of the generalized second law, and the
information loss puzzle.  Black hole entropy and the generalized second
law were introduced in  1972.\reftorange{Bek72}{Bek73}{Bek74}  A lot of
activity in black hole thermodynamics followed Hawking's 1974--75 papers
describing the Hawking radiance.\refto{Hawk74,Hawk75}  The information puzzle
dates from Hawking's 1976 paper.\refto{Hawk76}  Interest in these matters
mellowed at the end of that decade.   From the early 1990's there has been a
intense resurgence of interest in all three issues leading to much debate,
illumination and confusion.  Today, well into its third decade of development,
black hole thermodynamics remains intellectually stimulating and puzzling
at once.  What follows is not so much a full review of the first two issues, as
my impression of some promising directions which are likely to influence
resolution of the information puzzle and lead to insights outside the immediate
subject.

Black hole entropy had some predecessors: Christodoulou's irreducible
mass,\refto{Chris} Wheeler's suggestion of a demon who violates the second law
with help of a black hole,\refto{BekPT} Penrose and Floyd's observation
that the event horizon area tends to grow\refto{PenrFloyd} and Hawking's area
theorem.\refto{HawkArea} Carter\refto{Carter} and Bardeen, Carter and
Hawking\refto{BCH} were aware of the analogy between horizon area and entropy
as reflected in their first and second laws of black hole mechanics, but did
not take the analogy seriously.  The view that horizon area  divided by
Planck's length square is really an entropy, not just an analog of
entropy,\reftorange{Bek72}{Bek73}{Bek74} met initially with
opposition\refto{BCH,BekPT,Israel87} but  was embraced
\eject
\noindent widely after Hawking's demonstration\refto{Hawk75} that black holes
radiate thermally.   By the end of the 1970's it was generally accepted that a
black hole, at least a quasistatically and semiclassically evolving one, is
endowed with an entropy  (throughout I use units with $G=c=k$, but display
$\hbar$)             $$
S_{_{BH}}=A/ (4\hbar)  \eqno(SBH)
$$
Today it is clear that if one sticks to general relativity or to dilaton type
gravity theories  in $3+1$ dimensions, and matter has normal properties,
\Eq{SBH} is widely valid.\refto{Kallosh}

As a geometric property, black hole entropy could be granted thermodynamic
status only because of two points.  First, one can derive from it a
temperature by the thermodynamic relation $T=(\partial M/\partial S_{_{BH}})$
with $M$ the black hole mass--energy\refto{Bek73} which happens to have the
same form as Hawking's radiance temperature [$T_{_{BH}}=\hbar/(8\pi M)$ for
Schwarzschild]; in fact this is the way the proportionality contant in
\Eq{SBH} was first calibrated.\refto{Hawk74,Hawk75}  However, the
ulterior meaning of black hole entropy has remained a mystery.  Second, black
hole entropy enters into the generalized second of thermodynamics (GSL) on the
same footing as ordinary matter--radiation entropy $S_{\rm rad+mat}$: for a
transformation of a closed system including black holes
$$
  \Delta S_{_{BH}} + \Delta S_{\rm rad+mat} \geq 0  \eqno(GSL)
$$
This law has proved quite succesful.  Suffice it to recall that when it
was originally formulated,\refto{Bek72,Bek73} Hawking's radiance
was still a thing of the future, yet the GSL was found to be satisfied by the
Hawking process (in its semiclassical form).\refto{Bek75,Hawklet}  Since then
a number of succesful tests of the GSL have been carried out.  Two questions
arose: what  mechanism insures that the generalized entropy grows in any
situation, and are there any exceptions to the law ?  These are not trivial
questions: understanding why the {\it ordinary\/} second law (with no black
holes) works in the quantum world is just beginning to crystallize a century
and a half after Carnot, Clausius and Kelvin (see \Ref{Sorkin86} for a nice
recap).

The Hawking ``evaporation'' of a black hole brings with it the information
puzzle.\refto{Hawk76}  Recall the essentials.  Hawking's original
derivation and subsequent work show the radiance to have a thermal character
(quasi--Planckian spectrum and thermal statistics
mode--by--mode).\refto{ParkWald,Bek75}  This is usually traced to the picture
of pair formation out of the vacuum for modes that skim the event horizon on
their way out to future null--infinity ${\cal J_{+}}$.  One of each pair goes
out to contribute to the Hawking radiance; its companion is lost down the black
hole.  The quantum state of the Hawking radiation {\it by itself\/} lacks the
quantum correlations with the ``lost'' quanta which are part and parcel of the
original pure vacuum state at past null--infinity ${\cal J_{-}}$.  Hence the
Hawking radiation all by itself is in a mixed state.  It happens to be a nearly
maximally mixed state, and so is thermal. If the black hole truly disappears by
evaporation, one is left with a thermal (mixed) state of
radiation with nothing to correlate with in order to reconstitute the pure
state.  Hawking concluded from this that black hole evaporation catalyzes
unitarity violation, that quantum mechanics is not fully correct in the
presence of black hole horizons, and that contrary to the venerable rules, a
pure state can become mixed.\refto{Hawk76}  This strong claim forms the basis
of the information puzzle or  paradox in black hole physics.

The three issues are actually one in the sense that when people find out
how to fundamentally resolve one of them, they will have resolved all three.
In the last few years it has been fashionable to explore these issues in the
framework of  exactly solvable field--theoretic toy  models in $1+1$
dimensions.\refto{CGHS}  I wish my comments to be interpreted in unfashionable
$3+1$ dimensions.  What is lost in exactness of treatment this way is balanced
by the realism of the conclusions.

\vglue 0.6cm
\leftline{\twelvebf 2. The Meaning of Black Hole Entropy}
\vglue 0.4cm

``Entropy'' must be one of the most abused terms in physics.  We all agree
that Boltzmann's entropy derived from the one--particle distribution function
of a gas, and Gibbs' canonical ensemble entropy are closely related to
Clausius's thermodynamic entropy.  Somewhat more removed, but
still clearly related to phenomenological entropy, is Shannon
entropy\refto{Shannon} -- the measure of unavailable information,
$$
S=-\sum_A p_A \ln p_A \eqno(Shannon).
$$
Most likely unrelated to it are Kolmogorov entropy in the theory of chaotic
flows, or Chaitin's algorithmic entropy in the theory of computation.

Although there can be little doubt that black hole entropy  corresponds
closely to a phenomenological entropy, its deeper {\it meaning\/} has remained
mysterious.    Is it  similar to that of ordinary entropy, {\it
i.e.\/} the logarithm of a count of internal black hole states associated with
a
single black hole exterior ?\refto{Bek73,Bek75,Hawklet}  Is it the logarithm of
the number of ways in which the black hole might be formed
?\refto{Bek73,Hawklet}    Is it the logarithm of the number of horizon
quantum states ?\reftorange{Wheeler}{tHooft90}{STU}  Does it stand for
information lost in the transcendence of the hallowed principle of unitary
evolution ?\refto{Hawk76,Giddings94}   I would claim that at this stage the
usefulness of any proposed interpretation of black hole entropy turns on
how well
it relates to the original ``statistical'' aspect of entropy as a measure of
disorder, missing information, multiplicity of microstates compatible with a
given macrostate, {\it etc.\/}

In Hawking's field theoretic approach, which served as model for nearly
all work in the 1970's and 1980's, and in the venerable
surface gravity method,\refto{BCH,Visser1} black hole temperature is the
primary
quantity, and the black hole entropy is recovered from Clausius's rule $S=\int
dM/T$.  The statistical aspect is not exposed.  Wald's N\"other
charge method,\refto{WaldNoether,JacobsonKang} the method of deficit
angle,\refto{Banados} and the method of field redefinition\refto{JacobsonKang}
are likewise good for calculating black hole entropy in unfamiliar situations,
but leave one mostly in the dark as to its statistical meaning.  In the
Gibbons--Hawking Euclidean method\refto{GibbHawk77} the black hole entropy is
basically classical: the ${A/(4\hbar)}$ contribution appears at tree
level, {\it i.e.,\/} to lowest order in $\hbar$ in the functional integral.
Yet in statistical mechanics of fields, statistical entropy first appears at
one--loop level.  Thus although the Gibbons--Hawking approach has proved
fruitful for calculating the value of the black hole entropy in novel
situations,\refto{Kallosh}  it is not in itself a statistical interpretation
of black hole entropy (my early reaction to the Gibbons--Hawking approach is
recorded in the discussion to \Ref{Bek79}).  One might expect that going
beyond tree level might bring in truly statistical features of entropy.  Thus
enters entanglement entropy.

\vglue 0.4cm
\leftline{\twelveit 2.1. Why Entanglement Entropy ?}
\vglue 0.3cm

Entanglement entropy was  used very early in relativity to understand the Unruh
effect as resulting from ignoring the states  beyond the Rindler
horizon.\refto{Dowker,Israel2}  The last year witnessed a renaissance of the
interpretation of black hole entropy in terms of quantum entanglement entropy
proposed by Bombelli, Koul, Lee and Sorkin\refto{BKLS} (henceforth BKLS) in a
classic paper from the quiet period of the subjects's history. The idea was
rediscovered by Srednicki\refto{Srednicki} who pointed out that the global
vacuum state of a scalar field in flat spacetime, when restricted to the
exterior of an imaginary sphere, is in a mixed state there.  The  density
matrix of this mixed state arises from tracing out those parts of the global
state that reside inside the sphere; its entropy is evidently related to the
unknown information about the sphere's interior.  This entropy is nonvanishing
only because the exterior state is correlated with the interior one (or
entangled in the sense that the parts of the singlet state of two electrons
$|\,\uparrow\rangle|\,\downarrow\rangle-|\,\downarrow\rangle|\,\uparrow\rangle$
are entangled).
In the sphere's case the quantum entanglement entropy comes out
to be proportional to the sphere's surface area, albeit with a coefficient
which diverges quadratically in the high frequency cutoff.\refto{Srednicki}

The main points had been made earlier by  BKLS.  They also gave reasons
for relating at least part of the black hole entropy to  entanglement entropy
of the state outside black hole.  In particular, they pointed out that whereas
for an ordinary ``black box'' situation  the emergence of entanglement entropy
out of a pure state is to a large extent a matter of choice for the observer,
for the black hole case the horizon's presence makes its emergence mandatory.
They noted that because the black hole exterior evolves autonomously -- no
information is fed into it from inside the horizon --  one can expect a second
law to apply to an entropy defined exclusively in it.  BKLS were  aware that
the  divergence of the entanglement entropy is
due to high frequency modes near the
horizon, and suggested that the physical entropy is finite due to quantum
fluctuations of the geometry at the horizon.

Entanglement entropy has lately been
explored further by Susskind, Thorlacius and
Uglum\refto{STU} with an eye on the relation between
entanglement and radiation entropy.  Holzhey\refto{Holzhey} and Callan and
Wilczek\refto{CalWil}  have made use of clever techniques for computing it,
concluding with BKLS and Srednicki that a plane boundary in Minkowski
spacetime,
when the quantum state beyond it is ignored, gets ascribed entanglement entropy
proportional to the area of the boundary with an ultraviolet quadratically
divergent coefficient.
Kabat and Strassler\refto{KabStras}  further show that the
density operator in
question is thermal irrespective of the nature of the field.
Holzhey, Larsen and Wilczek\refto{HolLarWil} explore a method to regularize the
divergence in conformal field theories.

\vglue 0.6cm
\leftline{\twelveit 2.2. Entanglement Entropy is Operationally Finite}
\vglue 0.4cm

The divergence of entanglement entropy, common to flat and black
hole  spacetimes, has puzzled people.  But, at least in flat
spacetime, the problem is a red herring: when the operational procedure behind
the formal ``tracing'' is a physical one, there cannot be a divergence.  To see
why this is so, let me first state the problem as usually conceived.  The
global
vacuum state will be denoted by $|0\rangle$.  The space is divided by a
boundary into an interior and
exterior region.   Let a complete basis of states for
the interior region be
denoted by $\{|a\rangle\}$ and one for the exterior one by
$\{|A\rangle\}$.
Now suppose that $\{|a\rangle\}\otimes\{|A\rangle\}$ is a basis
for the global states.  Then it is possible to represent the vacuum state as
$$
|0\rangle=\sum_{aA} C_{aA} |a\rangle \otimes |A\rangle
\eqno(vacuumvector)
$$
where the $C_{aA}$ are complex numbers.  If nothing is known
about the interior side of the boundary, then one obtains the exterior state by
assigning each interior state  an equal weight, {\it i.e.,\/} by tracing
$|0\rangle\langle 0|$ over the basis  $\{|a\rangle\}$ and then normalizing:
$$
\hat r_{\rm ext}={{\rm Tr}_a\ |0\rangle\langle 0|\over {\rm Tr}_{aA}\
|0\rangle\langle 0|} ={\sum_{AB}\sum_a C_{aA}\,C^*_{aB}\  |A\rangle \langle
B|\over \sum_{A}\sum_a |C_{aA}|^2} \eqno(rhoout)
$$
It is the von Neumann entropy  $S =  -{\rm Tr}_A\ \hat r_{\rm
ext}\ln \hat r_{\rm ext}$ which is the entanglement entropy.  It diverges
because there are many high frequency modes in the sum in \Eq{vacuumvector},
and thus an infinity of states are traced over.

However, the kind of trace in \Eq{rhoout} does {\it not\/} correspond to
any operational prescription.  It is untrue, in general, that one knows nothing
about the interior state.  For example, if the region selected is spherical of
radius $R$, then just from the fact that the spacetime is flat to some
accuracy,
one knows that the energy $E$ associated with the interior region  has to be
small.  Of course, the global vacuum has zero energy, but one is discussing the
energy of the state left after tracing -- a different one.  In fact if the
boundary delineating the region being traced out were absolutely sharp, the
uncertainty principle might suggest a very large energy for it.  Thus we think
of that boundary as slightly fuzzy.

Anyway, we can write $E/R = \xi \ll 1$ where $\xi$ is the relativistic quality
parameter ($GM/c^2R$ in dimensional notation).  Thus in forming the density
operator for the exterior region, one should assign equal nonvanishing weights
only to the interior states with energy below $E$.  Equivalently, one should
trace  $|0\rangle\langle 0|\Theta(E-\hat H_{\rm int})$ instead of just
$|0\rangle\langle 0|$ over $\{|a\rangle\}$; here $\hat H_{\rm int}$ is the
Hamiltonian for the interior degrees of freedom.  In the expression for the new
density operator, $\hat\rho_{\rm ext}(E)$, all sums over $a$  are to be
confined
to states $|a\rangle$ with energy below $E$.  The claim is that the physical
entropy   $S_{\rm ext}(E)=-{\rm Tr}_A\ \hat\rho_{\rm ext}(E)\ln \hat\rho_{\rm
ext}(E)$ is finite for bounded $E$.

To see this most easily suppose the basis $\{|A\rangle\}$ diagonalizes
$\hat\rho_{\rm ext}(E)$.  Then the eigenvalues of $\hat\rho_{\rm ext}(E)$,
$$
p_A(E)={\sum'_a |C_{aA}|^2\over \sum_A\sum'_a |C_{aA}|^2}
\eqno(probability)
$$
are the probabilities for the exterior states $\{|A\rangle\}$; here a prime on
a sum means it is restricted to states with energy below $E$. Then the von
Neumann entropy of $\hat\rho_{\rm ext}(E)$ is just the Shannon entropy:
$$
S_{\rm ext}(E)=-\sum_A p_A(E) \ln p_A(E) \eqno(exterior)
$$
I now argue that this entropy is bounded from above by $\ln N(E)$, where
$N(E)$ is the number of {\it interior} quantum states $|a\rangle$ with energy
below $E$, itself a number easy to bound.

The first step is the well known symmetry theorem\refto{vonNeumann} whose proof
in the present context goes as follows (see \Ref{Srednicki}).   Define
$\hat\rho_{\rm int}$ by the analog of \Eq{rhoout}, but with the trace taken
over
$\{|A\rangle\}$.  This is the state of the interior region when one ignores the
information about the exterior. However, as before, interior states $|a\rangle$
with energies above $E$ are not allowed.  Thus the rows of the matrix $C_{aA}$
corresponding to such states are to be amputated in a physical discussion.  In
effect, given
the information that the interior region has little energy, the full
$C_{aA}$ does not give the correct global quantum state compatible with that
information.   Call the amputated matrix ${\cal C}$.  Then  the manifestly
positive definite  matrix $R_{\rm ext}\equiv {\cal C}^\dagger{\cal C}/{\rm
Tr}\,
{\cal C}^\dagger{\cal C}$
represents $\hat\rho_{\rm ext}$ (see \Eq{rhoout}) while
$R_{\rm int}={\cal C}^*{\cal C}^{{^T}}/{\rm Tr}\,{\cal C}^*{\cal C}^{{^T}}$
represents $\hat\rho_{\rm int}$ ($T$ denotes ``transpose'').

Because the sets  $\{|A\rangle\}$ and $\{|a\rangle\}$ are inequivalent,
$R_{\rm int}\not=R_{\rm ext}$. However, ${\rm Tr}\,{\cal C}^*{\cal
C}^{{^T}}={\rm Tr}\,{\cal C}^\dagger{\cal C}$ (transposing does not affect
traces). By the cyclic invariance of the trace of a product, it is easy to
extend this to  ${\rm Tr}\,({\cal C}^*{\cal C}^{{^T}})^n={\rm Tr}\,({\cal
C}^\dagger{\cal C})^n$ for $n=2, 3,\dots\quad$  Equivalently, $\sum_A p_A^n=
\sum_a p_a^n$ where $p_a$ is an eigenvalue of $\hat\rho_{\rm int}$ defined in
analogy with \Eq{probability}.  This last relation is true for all $n$ only if
$R_{\rm int}$ and $R_{\rm ext}$ have the same list of nonvanishing
eigenvalues (the number of zero eigenvalues may be
different\refto{DanSchiff}).  Now because the von Neumann entropy of
$\hat\rho_{\rm int}$, $S_{\rm int}$, can be expressed in terms of $p_a$ in
analogy with \Eq{exterior},
$$
S_{\rm ext}(E)=S_{\rm int}(E)=-\sum_a{}'\ p_a \ln p_a   \eqno(Sineq)
$$
Of course this key result would likewise  be valid formally had one not
excluded
the high energy $|a\rangle$ states.  However, since $S=S_{\rm
ext}(\infty)=\infty$, that result would  not be interesting.

The maximum possible value of $S_{\rm int}(E)$ is obtained when all $p_a$
are equal.  If there are $N(E)$ interior $|a\rangle$ states below energy $E$,
then the sum in \Eq{Sineq} equals $\ln N(E)$, the microcanonical entropy of
the interior
as a function of energy.  Thus one finds for the entanglement entropy
according to the exterior observer
$$
S_{\rm ext}(E)<\ln N(E)  \eqno(inequality)
$$

The terms of the problem require that the states $|a\rangle$ counted by $N(E)$
be confined to the
interior region.  One way to enforce this is to subject the field
to a boundary condition at the surface between the regions, which amounts to
putting the system represented by $\hat H_{\rm int}$ in a box and ignoring the
exterior. Suppose the field is free, and thus described by some one--particle
Hamiltonian $\hat h_{_1}$.
Then a semianalytical argument\refto{Bek84} assures one
that for any box shape
$$
{\rm max}\left[\ln N(E)/E\right] \approx [\zeta(\hat h_{_1},
4)]^{1/4}\eqno(zetaformula)
$$
where
$$
\zeta(\hat h_{_{1}}, 4) \equiv {\rm Tr}\ \hat h_{_{1}}^{-4}=
g_{_{1}}\ \varepsilon_{_{1}}^{-4} + g_{_{2}}\ \varepsilon_{_{2}}^{-4}+\dots
\eqno(zeta)
$$
is the analog of Riemann's zeta function $\zeta(4)$, but with the
one--particle eigenenergies $\varepsilon_{_j}$ (each with multiplicity
$g_{_j}$)
in the box replacing the positive integers.   This result has been
checked\refto{Bek84} by counting all many--particle states in a box up to
energy
$E$ for electromagnetic, scalar and neutrino fields.  The boxes were either
spherical, or rectangular with various aspect ratios.  The boundary conditions
were Neumann or Dirichlet for the scalar, conducting boundary for the
electromagnetic, or zero energy outflow for the neutrino field.  The results
confirm \Eq{zetaformula} to about 5\% accuracy.  It is already plain from the
comparison of
inequality \(inequality) with the approximation \(zetaformula)  that
the entanglement entropy arising from ignoring the interior region is bounded
so
long as it is recognized that the interior region has limited energy $E$.  The
entanglement entropy grows at most as fast as $E$.

The approximation \(zetaformula) can be traded for the {\it rigorous\/}
bound\refto{BekSchiff89, BekSchiff90}
$$
\ln N(E) < [4!\,\bar\zeta(\hat h_{_{1}}, 4)]^{1/4} E, \eqno(newzetaformula)
$$
where a bar indicates that the eigenenergies and degeneracies used to calculate
the zeta function are to be those appropriate for a sphere with radius $R$
equal to the circumscribing radius of the box (which can be of any shape and
topology). Now the terms in \Eq{zeta} typically drop off rapidly.
And since only $[\bar\zeta(\hat h_{_{1}}, 4)]^{1/4}$ is of concern, and
$g_{_1}$
should not be large compared to unity, a passable approximation to
$[\bar\zeta(\hat h_{_{1}}, 4)]^{1/4}$ is $1/\varepsilon_{_1}$.  On dimensional
grounds one expects,
for a massless field, that $\varepsilon_{_1}\sim \hbar/R$.  If
the field is massive, $\varepsilon_{_1}$ should be larger.  Thus for a massless
field one expects $[\bar\zeta(\hat h_{_{1}}, 4)]^{1/4} \approx R/\hbar$, with a
smaller value for a massive field.   Explicit calculation of $\bar\zeta(\hat
h_{_{1}}, 4)$ for electromagnetic, scalar and neutrino
fields\refto{Bek81,BekSchiff89}
confirm this.  One can cover every type of known
field by replacing \(newzetaformula) by the (rather generous) uniform bound
$$
\ln N(E) < 2\pi RE/\hbar  \eqno(mybound)
$$
which I like to  call the universal entropy bound.\refto{Bek81}

Put all this together.  From the definition of relativistic quality
parameter one has $E=\xi R$.  Substitution in bound \(mybound) and that in
inequality \(inequality) gives
$$
S_{\rm ext} < 2\pi\xi R^2/\hbar   \eqno(finite)
$$
which is the desired formula.  This bound on entanglement entropy scales up
with
area of the circumscribing sphere, but the coefficient is not infinite  (for a
nearly flat spacetime system, $\xi\ll 1$).  Let me now cavalierly push the
formula beyond its intent to $\xi\rightarrow 1$ (the
black hole regime).  Obviously the entanglement entropy could very well
approach $\pi R^2/\hbar$ which is of the  order of the black hole entropy,
\Eq{SBH}  The identification of the two\refto{Srednicki}  seems reasonable on
this grounds.

An obvious caveat about the above argument is that it pushes bound \(mybound),
which is well established in flat spacetime, to a strong gravity situation.
The strong gravitational redshift in the black hole vicinity may well allow
many states based on arbitrarily high (local) frequency modes to be included in
the count of states for a system with finite global energy $E$ (isn't this what
the Hawking process is about ?).  Thus, although it is clear than in flat
spacetime the entanglement entropy is finite in physically well defined
situations, the analogous claim about curved spacetime awaits proof of the
analog of bound \(mybound) for strong gravity.   A quite independent
argument that black hole entropy calculated as entanglement
entropy will come out finite as a result of renormalization of the
gravitational constant is put forth by Susskind and Uglum.\refto{SusUglum}

\vglue 0.4cm
\leftline{\twelveit 2.3. The Multiplicity of Species Problem}
\vglue 0.3cm

Another thorny problem with equating entanglement and black hole entropy
is that since each field in nature must make its contribution to
the entanglement entropy,  black hole entropy should scale up with the number
of
field species in nature.
Yet \Eq{SBH} says nothing about number of species !  An
interesting resolution suggested by Sorkin\refto{Sorkin83} and 't
Hooft\refto{tHooft93} is that indeed different species contribute, but that
the contributions of the actual species in nature exactly add up to
$A/(4\hbar)$.  The point of view here is that the list of elementary particle
species is prearranged to chime with gravitational physics.  The results of
Sec.~2.2 can be used to show that this is not out of the question.  One adds up
the specific values of $\bar\zeta(\hat h_{_{1}}, 4)$ for the species found in
nature to form a grand zeta function for ``matter''.  Taking into account
three  species of single--helicity neutrinos, six species of quarks, three of
leptons and eight gluons together with all their antiparticles, as well as
the photon, the $W^\pm$ and $Z$ bosons, and a Higgs doublet of complex
scalars (for simplicity I think of all species as massless), one
gets\refto{Bek84,BekSchiff89}  $\bar\zeta({\rm grand}, 4)=9.45R^4/\hbar^4$.
Repeating the argument based on inequality \(newzetaformula), one is led to
replace inequality
\(finite) by $S_{\rm ext} < 3.88\xi R^2/\hbar$.  Thus it is not
inconceivable
that due to the gravitational and other interactions, $S_{\rm ext}$
ends up being  $\pi R^2/\hbar$ in the strong gravity limit, as appropriate for
black hole entropy.

A very different resolution is offered by Jacobson.\refto{Jacobson} The
argument is, roughly, that the effective action of every field quantized in
curved spacetime carries a piece that looks like the Hilbert action, so that
every such field makes a correction to the value of $G^{-1}$.  For $n$ fields
the correction to $G^{-1}$ is proportional to $n$.  But the entanglement
entropy contributed by $n$ fields is also proportional to $n$.   Thus, if all
of $G^{-1}$ comes from effective actions (Sakharov's vision of effective
gravity) the entanglement entropy will scale up as $G^{-1}$.  It thus makes
sense to identify the entanglement entropy and the black hole entropy; the
latter, $c^3A/(4G\hbar)$ in dimensional form, also scales  like $G^{-1}$.
A similar argument is given by Susskind and Uglum.\refto{SusUglum}
This resolution of the multiplicity problem depends on black hole
entropy being all entanglement entropy.  As I argue below, this
identification  seems to resolve the information loss puzzle in a somewhat
too trivial way.

This section would be incomplete without reference to the resolution due to
Frolov.\refto{Frolov}  Its background is Frolov and
Novikov's\refto{FrolovNovikov} identification of black hole entropy  with the
entanglement entropy of the mixed state obtained by tracing over the {\it
exterior\/} states in the global vacuum of a field.  Note that it is a logical
consequence of the  nature of quantum entanglement that one cannot have the
black hole entropy residing in one region and arising from ignorance of the
state of the degrees of freedom in that same region.  Accordingly, Frolov and
Novikov's black hole entropy ``resides'' inside the black hole.  The BKLS and
Frolov--Novikov viewpoints do not necessarily clash because the symmetry
theorem certifies that, because the global state is pure, the entanglement
entropy comes out the same either way.

Frolov worried about the dependence of this entanglement entropy on the number
of matter fields, and reconsidered the identification. He recalls that the
free energy ${\cal F}$
of a system depends on an external parameter $\lambda$ via,
say, the dependence of mode frequencies on it.  Thus $d{\cal F}=- S dT + \Pi
d\lambda$. Here $S$ is the usual statistical entropy, and the extra term is
usually interpreted as work.  Frolov regards black hole temperature as an
external parameter, at least within York's picture\refto{York} of the black
hole enclosed in a cavity whose wall is kept a fixed temperature.  Since mode
frequencies scale inversely with black hole mass, Frolov considers them as
proportional to $T_{_{BH}}$; then $d\lambda\propto dT_{_{BH}}$ and Frolov
interprets the entire coefficient of $-dT_{_{BH}}$ in $d{\cal F}$, not just
$S$,
as $S_{_{BH}}$.  He calculates that the new terms mostly cancel out the
entanglement entropy's contribution to $S$.  Since entanglement entropy is a
one--loop contribution, Frolov finds $S_{_{BH}}$ to be close to the
Gibbons--Hawking
tree--level entropy.  If little of the Hilbert action is induced
by quantum corrections,
this last entropy is independent of the number of species,
and so the species problem is resolved

I find Frolov's view of temperature somewhat confusing.  In addition, and on a
more practical level, I note that because of the negative specific
heat of the Schwarzschild black hole, it is possible for such a hole to be in
stable canonical ensemble with temperature as a parameter only in a very small
container.\refto{Hawklet,York}  What to do about black hole entropy for
a black hole in empty space or one in a large cavity ?  Mode frequencies of
radiation of a free black hole
are not functions of its temperature.  Thus at best,
Frolov's reinterpretation of the Frolov--Novikov paper is limited in scope.
However, as I discuss now, some reidentification of what is meant by black hole
entropy is indeed needed for another reason.

\vglue 0.4cm
\leftline{\twelveit 2.4. A Proposal for Black Hole Entropy}
\vglue 0.3cm

Consider a black hole formed from collapse of a classical object.  A (quantum)
scalar field, originally in the
vacuum state, propagates on this background.  On a
Cauchy hypersurface
like $\sigma\cup\upsilon_{_1}$ in the Penrose diagram of Fig.
1, the entanglement entropy of the state $\hat\rho_{\upsilon_{_1}}(\sigma)$
arising from tracing $|0\rangle\langle 0|$ over the interior quantum states on
$\upsilon_{_1}$ is, according to the BKLS viewpoint, just  $S_{_{BH}}$.  It
follows from the \bigskip \vbox{\hskip 40pt\hbox{\epsfbox{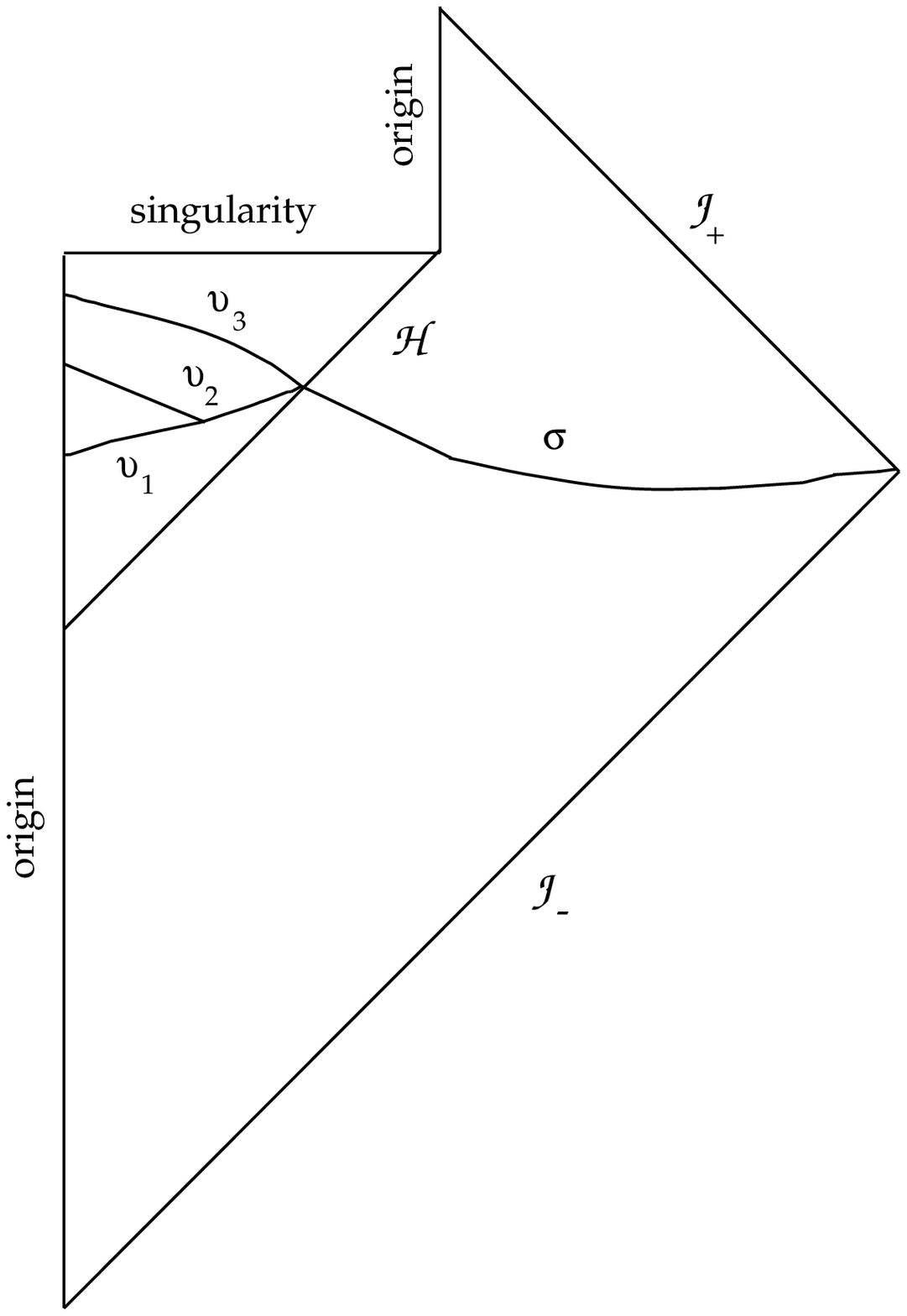}}}
\bigskip
{\rightskip=3pc
 \leftskip=3pc
\tenrm\baselineskip=12pt\noindent
Fig.1:  Penrose diagram for an evaporating Schwarzschild black hole showing the
semihypersurfaces $\upsilon_{_j}$ inside the horizon ${\cal H}$ and
$\sigma$ outside it.   \bigskip }

\noindent  symmetry
theorem that the {\it interior\/} entanglement entropy of the
state $\hat\rho_{\sigma}(\upsilon_{_1})$
which arises from tracing over the {\it
exterior\/}
states on $\sigma$  must also be equal to $S_{_{BH}}$ since the  global
state of the scalar field is pure.  But this ``interior'' entropy can be
identified with the  fine--grained entropy of the Hawking radiation since it
arises precisely from ignoring information about states in the black hole
exterior.
If the semiclassical picture is any guide, the horizon area will shrink
and thus $S_{_{BH}}$ must decrease as time goes on.  But then the radiation
entropy, which is at all
times equal to $S_{_{BH}}$ by the symmetry theorem,  must
decrease  and end up by vanishing as the black hole fizzles out. If this
conclusion is correct, it means that the radiation's state becomes fully pure
in
the limit.  This eventuality would obviously remove the information puzzle.

Yet despite venerable arguments in favor of such
an outcome,\refto{Page80,tHooft85} no sign of this returning of the radiation
to purity is seen either in the semiclassical\refto{Hawk76} or beyond
semiclassical\refto{Fiola} calculations.  This has engendered the thought that
the puzzle cannot even be properly stated without detailed understanding of
trans--Planckian physics.\refto{tHooft90,SusUglum}  There is thus an obvious
problem with the argument in the preceding paragraph.  I infer from this that
one should not rigidly equate entanglement entropy with black hole entropy. To
be sure, this is no new insight.  BKLS stated that entanglement entropy is only
a part of $S_{_{BH}}$.\refto{BKLS}
And Callan and Wilczek claimed that it  is only
a correction to the tree level part of $S_{_{BH}}$.\refto{CalWil}

Anyway, tracing over the interior states leaves open the question of which
semihypersurface $\upsilon$ this is being done on.  The exterior spacelike
semihypersurface $\sigma$ of interest (see Fig. 1) can be continued in any
number of ways -- $\upsilon_{_1}$, $\upsilon_{_2}$, $\upsilon_{_3}, \dots$ --
inside the horizon to the central point.  Tracing over the states on the
typical semihypersurface $\upsilon$ gives a density operator
$\hat\rho_{\upsilon}(\sigma)$ for the black hole exterior.  How does
$\hat\rho_{\upsilon}(\sigma)$ depend on the choice of $\upsilon$ ? Classically
it does not.  The global vacuum density operator $|0\rangle\langle 0|$ is
unevolving in the Heisenberg picture.  The interior observables do evolve in
that picture and so must their eigenstates.  However, their evolution from
$\upsilon_{_1}$ to $\upsilon_{_2}$, say, is unitary: no information is fed from
the exterior since the two spacelike semihypersurfaces meet at the horizon.
Thus if one performs the
trace of $|0\rangle\langle 0|$ over interior states in a
representation based on eigenstates of observables, one can expect the
resulting
density operator $\hat\rho_{\upsilon}(\sigma)$ to be the same for all choices
of  $\upsilon$ because the change from $\upsilon_{_1}$ to $\upsilon_{_2}$ is
equivalent to a change of basis, and traces are unaffected by  a change of
basis.  Thus, classically, $\hat\rho_{\upsilon}(\sigma)$ is unique for given
$\sigma$.

But quantum fluctuations of the geometry are bound to smear the
meeting point of the various $\upsilon_{_j}$ at the horizon (this
smearing is related to that invoked by BKLS to regularize the entanglement
entropy).
The unitary relation between interior eigenstates states on the various
semihypersurfaces $\upsilon$ cannot be relied upon because in the face of
the fluctuations the very meaning of the statement ``spacelike
semihypersurfaces
meet at the horizon'' becomes fuzzy.  My guess is that because of this
$\hat\rho_{\upsilon}(\sigma)$ depends slightly on $\upsilon$.

However, the entropies of the various $\hat\rho_{\upsilon}(\sigma)$,
namely
$S_{\upsilon_{_1}}(\sigma),  S_{\upsilon_{_2}}(\sigma), \dots$ {\it are\/}
identical.
By the symmetry theorem $S_{\upsilon_{_1}}(\sigma)$ equals the entropy
$S_{\sigma}(\upsilon_{_1})$ of the state induced on $\upsilon_{_1}$ by tracing
$|0\rangle\langle0|$ defined on $\sigma\cup\upsilon_{_1}$ over the states in
$\sigma$, and analogously for $\upsilon_{_2}, \upsilon_{_3}, \dots$\quad  Since
the global state $|0\rangle\langle 0|$
and the semihypersurface $\sigma$ are both
fixed, the trace, $\hat\rho_{\sigma}(\upsilon)$, and the corresponding entropy
$S_\sigma(\upsilon)$
have to be the same on all $\upsilon$. That and the symmetry
theorem gives
$S_{\upsilon_{_1}}(\sigma)= S_{\upsilon_{_2}}(\sigma)= \dots$\quad.

Since there are many possible, albeit quite similar, density operators
on $\sigma$, $S_{\upsilon}(\sigma)$
is not the full expression of the statistical
uncertainty on $\sigma$ about the black hole interior.  According to
information
theory,\refto{Shannon} if the states of a system can be classified into
several
classes $\{\kappa\}$, one gets the total uncertainty (entropy) as the sum of
the
expression $-\sum_\kappa p_\kappa \ln p_\kappa$, where $p_\kappa$ is the
probability of
class $\kappa$, and the weighted average of the intrinsic entropies
of the various classes (the weighing factors again being $p_\kappa$).
Obviously
in the present
case the $\kappa$ stand for the $\hat\rho_{\upsilon}(\sigma)$.  In
the spirit of Laplace's principle of ignorance, I shall assume that there are
effectively ${\cal N}$
equally likely $\hat\rho_{\upsilon}(\sigma)$, where ${\cal
N}$ is a finite number set by the amplitude of the quantum fuzziness alluded to
above.  Then       $$
{\rm Uncertainty\ on\ \sigma} =\ln {\cal N} + {\cal N}^{-1}\sum_j
S_{\upsilon_{_j}} (\sigma)=\ln {\cal N}+S_{\upsilon}(\sigma).
\eqno(split)
$$

Now $S_{\upsilon}(\sigma)=S_\sigma(\upsilon)$ in this expression is equal to
the Hawking radiation fine--grained entropy on  $\sigma$ ({\it
c.f.\/} argument at the beginning of this section).  I now interpret $\ln {\cal
N}$ as $S_{_{BH}}$
because it is the extra uncertainty about the state in the black
hole interior that is independent of the type of quantum state or field being
considered.  By construction this black hole entropy is independent of the
number of matter fields.  And in this interpretation the information puzzle is
not trivially removed: the eventual disappearance of  $\ln {\cal N}$ as the
horizon contracts does not force $S_\sigma(\upsilon)$ to vanish, though the
eventual ``purification'' of the Hawking radiation is certainly not forbidden.
It remains to be seen whether, because of quantum fluctuations, ${\cal N}$ is
indeed finite,  and whether its logarithm indeed scales as horizon area.  Since
within the interpretation just offered $\ln {\cal N}+S_{\upsilon}(\sigma)$ is
evidently the generalized entropy of \Eq{GSL}, a more immediate question is why
does this quantity tend to rise ?  In other words, what makes the GSL work ?

\vglue 0.6cm
\leftline  {\twelvebf  3. The Generalized Second Law at Work}
\vglue 0.4cm
\leftline{\twelveit 3.1. Early Arguments for the Validity of the GSL}
\vglue 0.3cm

Early ``proofs'' of the GSL used {\it gedankenexperiments\/} to show that a
loss of material entropy into a black hole is typically compensated by growing
$S_{_{BH}}$.\refto{Bek72,Bek73,Bek74}  With the advent of Hawking's radiance,
thermodynamic\refto{Hawklet,Sewell} and statistical\refto{Bek75,Page76}
arguments were given that any decrease in $S_{_{BH}}$ is more than offset by
the
growth of the radiance's entropy.  Hawking's argument\refto{Hawklet} is that
since the radiance is dumped into a low temperature enviroment,  the increase
in radiation entropy more than compensates for the reduction of entropy of the
hotter black hole.  Sewell\refto{Sewell} argued that the work done by a
system whose intensive parameters (temperature, electric potential)  are set by
a black hole should not, as in ordinary thermodynamics, exceed the reduction in
its Gibbs free energy.  By the conservation laws  this is equivalent to
requiring a growth in generalized entropy.   These arguments make it
seem that black hole thermodynamics is within the province of ordinary
thermodynamics;\refto{WaldErice} however, they leave one in the dark about the
statistical reasons for the GSL.

I demonstrated early\refto{Bek75} that the  statistics of the outgoing
Hawking radiance (also found in \Ref{ParkWald}) make it as entropic as
allowed by the spectrum that filters through the potential barrier around the
black hole; this remains true even when thermal radiation of any temperature
$T$ impinges on the hole.  The GSL is satisfied in both processes
mode--by--mode.   The processes of emission or reemission of incident
radiation are irreversible except when $T=T_{_{BH}}$, making it plain that the
radiation entropy studied is a coarse-grained one.  Page\refto{Page76} has
calculated that the Hawking radiance of a hole in free space carries 1.619
times more entropy than would be required to break even according to the GSL.
All the approaches mentioned so far {\it assume\/} that $T_{_{BH}}$ derives
from the horizon area, and do not explain why/whether the GSL always works
and how it fits in with quantum mechanics (which does not require an increase
in entropy).

\vglue 0.4cm
\leftline{\twelveit 3.2. Modern Proofs of the GSL}
\vglue 0.3cm

This last issue was first studied by Sorkin\refto{Sorkin86} in a seminal
paper, a jumping off point for the BKLS paper.  Sorkin ignores the black hole
interior, and assumes the exterior and horizon can be described by a density
operator $\hat\rho_{\rm ext}(\sigma)$ (here $\sigma$ is still an
exterior spacelike
semihypersurface).  He notes that $\hat\rho_{\rm ext}(\sigma)$
must evolve autonomously (not influenced by the goings on beyond the horizon),
at least in a classical picture of geometry.   Its properties of positive
definiteness, hermiticity  and unit trace are expected to be preserved by this
evolution.  Sorkin further assumes, on the ground of conservation of  energy
for the whole system, that the maximum possible value of $S_{\rm ext}(\sigma)$,
the von Neumann entropy  of $\hat\rho_{\rm ext}(\sigma)$, is unaffected by
evolution.  He then proves that all this leads to the growth of $S_{\rm
ext}(\sigma)$ as $\sigma$ is pushed forward in time.

Sorkin regarded
this an embryonic proof of the GSL, valid for dynamical black holes
as well as quasistatic ones.  He did not discuss how to split   $S_{\rm
ext}(\sigma)$ into black hole and radiation parts. It is clear from Sorkin's
characterization of  $\hat\rho_{\rm ext}(\sigma)$ that one may intuitively
identify $S_{\rm ext}(\sigma)$
with my ``uncertainty on $\sigma$''.  (But I see no
direct way to construct Sorkin's  $\hat\rho_{\rm ext}(\sigma)$ from my
$\hat\rho_{\upsilon}(\sigma)$). One thus gets  a natural split for $S_{\rm
ext}(\sigma)$, \Eq{split}.  Thus Sorkin's is a proof that
$\ln {\cal N}+S_{\upsilon}(\sigma)$ must increase as $\sigma$ advances.

A very different proof of the GSL for quasistatic changes of a black hole has
been formulated by Frolov and Page,\refto{FrolovPage} who were influenced by
Zurek and Thorne.\refto{Thorne} For an eternal black
hole, Frolov and Page consider the exterior mixed initial state $\hat\rho_{\rm
initial}$ to have a factorable form $\hat\rho_{\rm up}\otimes\hat\rho_{\rm
in}$,
where ``up'' denotes radiation modes coming up from the past horizon (in the
picture of an eternal black hole -- equivalent to Hawking radiation modes for
an
evaporating one), and ``in'' denotes modes ingoing from ${\cal J_{-}}$.  The
von
Neumann entropies are
thus related by $S_{\rm initial}= S_{\rm up}+ S_{\rm in}$.
The  state $\hat\rho_{\rm initial}$ is assumed to evolve unitarily  to a final
state $\hat\rho_{\rm final}$
so that $S_{\rm final}=S_{\rm initial}$.  The natural
modes for
this last are ``out'' modes escaping to ${\cal J_{+}}$ and ``down'' modes
falling into the future horizon.  By tracing $\hat\rho_{\rm final}$ over states
formed out of ``out'' modes they obtain $\hat\rho_{\rm down}$ and by tracing
out
``down'' type states they obtain $\hat\rho_{\rm out}$.  Because of correlations
between ``out'' and ``down'' quanta, $\hat\rho_{\rm final}$ is {\it not\/}
factorable as $\hat\rho_{\rm out}\otimes\hat\rho_{\rm down}$.  In fact the
correlations imply that  $S_{\rm final}< S_{\rm out}+ S_{\rm down}$.  Frolov
and
Page thus obtain
$$
\Delta S_{\rm rad+mat}= S_{\rm out} - S_{\rm in} >  S_{\rm
up} - S_{\rm down}     \eqno(change)
$$
In this approach no attempt is made to follow the entropy changes moment by
moment; only the overall change in ordinary entropy $\Delta S_{\rm rad+mat}$
is of import.

In terms of the energies measured at infinity,  the change in black hole
entropy is evidently $\Delta S_{_{BH}}= (E_{\rm in}(\infty)-E_{\rm
out}(\infty))/T_{_{BH}}$ .  By conservation of energy $E_{\rm in}(\infty)-
E_{\rm out}(\infty)=E_{\rm down}(\infty)-E_{\rm up}(\infty)$.  Converting
the energies to the frame of a local observer corotating near the horizon
(or at rest near it in the Schwarzschild case),  and using inequality
\(change), Frolov and Page are led to
$$
 \Delta S_{_{BH}}+\Delta S_{\rm
rad+mat} > [S_{\rm up} - E_{\rm up}({\rm local})/T_{_0}]  - [S_{\rm down}-
E_{\rm down}({\rm local})/T_{_0}]  \eqno(delta)
$$
where $T_{_0}$ is
$T_{_{BH}}$ blueshifted to the local observer's frame: $T_{_0}/T_{_{BH}}=
E({\rm
local})/E(\infty)$.  Frolov and Page regard the ``up'' and ``down'' systems
as strictly equivalent by time reversal invariance.  The ``up'' states
coming out of  the past horizon are supposed to be in equilibrium at global
temperature $T_{_{BH}}$ and thus at $T_{_0}$ in the local observer's frame.
The ``down'' modes form the same system, but in some other state.  Frolov
and Page recall that when $S$ and $E$ are properties of a thermodynamic
system in any state, and $T_{_0}$ is some fixed temperature, $S-E/T_{_0}$
attains its maximum when the system is in equilibrium at temperature $T_{_0}$.
Thus, conclude Frolov and Page, the r.h.s. of inequality \(delta) must be
positive, and  the GSL \(GSL) follows.

How general is the Frolov--Page proof of the GSL ?  It is, of course, limited
by its reliance on the semiclassical approximation (classical geometry driven
by averages of quantum stress tensor).   This weakness is remediable.  Fiola,
Preskill, Strominger and Trivedi\refto{Fiola} have recently  devised a proof
of the GSL in $1+1$ dimension dilaton gravity which goes beyond semiclassical
considerations.  However, that proof is restricted to very special
situations, and works only if a new type of entropy is ascribed
to coherent radiation states.  Frolov and Page's proof certainly has a wider
applicability.  But it does have a loophole: the assumed equivalence of ``up''
and ``down'' systems by time reversal invariance.  The eternal black hole
(Kruskal spacetime) is time--reversal invariant as assumed; the realistic
radiating black hole is not (a time reverted black hole is not a black hole).
Can one project this equivalence of systems from the former to the later ?

What is involved in the statement that $S-E/T_{_0}$ is maximum at
equilibrium at temperature $T_{_0}$ ?  The state of the matter
and radiation is encoded in some a density operator $\hat\rho$.  In terms of
the hamiltonian $\hat H$, $E={\rm Tr}(\hat\rho \hat H)$ while $S=- {\rm
Tr}(\hat\rho\ln\hat\rho)$.
Thus the variation  $\delta(S-T_{_0} E)$ under a small
variation $\delta\hat\rho$ is
$$
\delta(S-T_{_0} E)={\rm Tr}[\delta\hat\rho(\hat H+T_{_0}\ln
\hat\rho+T_{_0})]   \eqno(extremum)
$$
so that $S-T_{_0} E$ has an extremum under variations that preserve
${\rm Tr}\,\hat\rho=1$ where $\hat\rho$ satisfies  $\hat H+T_{_0}\ln
\hat\rho+T_{_0} -\lambda=0$ with $\lambda$ a Lagrange multiplier.  Obviously
there is a unique solution $\hat\rho\propto\exp(-\hat H/T_{_0})$, {\it i.e.,\/}
there is one extremum of $S-T_{_0} E$, a thermal (equilibrium) state with
temperature $T_{_0}$.  This extremum is a maximum since for fixed $E$, $S$
attains a maximum in equilibrium.   Thus the r.h.s. of \Eq{delta} is
indeed nonnegative provided the ``up'' and ``down'' states are described by
the selfsame hamiltonian.

For the eternal black hole time reversal invariance does indeed guarantee
equivalence of ``up'' and ``down'' hamiltonians.  Compare now an
evaporating black hole made by collapse with an eternal black hole of like
parameters.  Assuming a complete set of states, each of the relevant
hamiltonians can be expanded in the usual form $\hat H=\sum |j\rangle\langle
j|\epsilon_{_j}$.  If the time variation of the evaporating black
hole's  parameters may be ignored, the ``down'' states and eigenenergies for
the
two black holes
are in detailed correspondance, so that the ``down'' hamiltonians
are equivalent.  Thus the ``down'' hamiltonian for the evaporating black hole
is
equivalent to the ``up'' hamiltonian of the eternal black hole.  But the
equivalence cannot
be carried one step further.  The Hawking ``up'' states from an
evaporating black hole emerge through the {\it time dependent\/} geometry of
the
collapsing object.
Thus they cannot be put into exact correspondance with ``up''
states for the eternal black hole which emerge right into the stationary
geometry.  This is particularly true of early emerging ``up'' states.  Thus the
exact equivalence of ``up'' and ``down'' hamiltonians for the realistic
evaporating black
hole is in question since the comparison must be over a complete
set of states.

The above mathematical nicety may well prove irrelevant for the Frolov--Page
proof when it is the scattering of microscopic systems off the black hole which
is under consideration.  However, for events involving an evaporating black
hole and macroscopic objects, the sets of ``up'' and ``down'' modes are
distinctly different.  Macroscopic objects are bound states of many quanta of
elementary fields.  As discussed below, over the black hole evaporation
lifetime such an object occurs in the Hawking radiance only with exponentially
small probability.  Thus even if emitted, the object is emitted by a black
hole whose parameters cannot be regarded as stationary  even in rough
approximation.  The comparison of the realistic and eternal black holes is
thus murky since the former evolves drastically over the relevant time span.
The equivalence of the ``up'' and ``down'' hamiltonians is thus unclear, and
inequality \(delta) cannot be exploited.

\vglue 0.4cm
\leftline{\twelveit 3.3. The Universal Entropy Bound from the GSL}
\vglue 0.3cm

As just mentioned, the Frolov--Page proof is unconvincing for a situation
where macroscopic matter falls into a black hole.  Such a situation occurs
frequently, {\it e.g.,\/} astrophysical accretion onto a black hole.  The
Sorkin proof does seem to apply.  Thus I assume that the
GSL is also valid in such a situation.  There are then interesting
consequences.

First  consider dropping a spherical macroscopic system of mass $E$, radius
$R$ and entropy $s$ into a Schwarzschild black hole of mass  $M\gg E$ from a
large distance $D\gg M$ away.  The black hole gains mass $E$, which it then
proceeds to radiate over time $\tau$.  At the end of this process the black
hole is back at mass $M$.  Were the emission reversible, the radiated entropy
would be
$E/T_{_{BH}}$.  As mentioned, the emission is actually irreversible, and
the entropy emitted
is a factor $\mu>1$ larger.  Thus the overall
change in generalized entropy is
$$
\Delta S_{_{BH}}+ \Delta S_{\rm rad} = \mu E/T_{_{BH}} - s   \eqno(netchange)
$$
{}From numerical work Page\refto{Page76} estimates $\mu=1.35-1.64$ depending on
the species radiated.   One can certainly  choose  $M$ larger than $R$ by an
order of magnitude, say, so that the system will fall into the hole without
being torn up: $M=\gamma R$ with $\gamma = {\rm a\ few}$. Thus, if the GSL is
obeyed,  the restriction
$$
s < 8\mu\gamma\pi RE/\hbar   \eqno(newbound)
$$
must be valid.   It is clear from the argument that there is no need for
$\mu\gamma$ to be arbitrarily large.  Thus from the GSL one infers  a bound on
the entropy of a rather arbitrary -- but not strongly gravitating -- system in
terms of its total gravitating energy and size.  Note that this bound
is compatible with
bound \(mybound) which comes from statistical mechanics in flat
spacetime.

One objection that could be raised to the above line of argument is that
Hawking radiation pressure might keep the system from being absorbed by the
hole, thus obviating the conclusion.  This is not so. Approximate the
Hawking radiance as black body radiance of temperature $\hbar/(8\pi M)$ from a
sphere of radius $2M$, the energy flux at distance $r$ from the hole is
$$
F(r)= {\hbar\over 61,440(\pi M r)^2}   \eqno(flux)
$$
resulting in a radiation force $f_{\rm rad}(r)=\pi R^2 F(r)$ on the infalling
sphere.  Writing the  Newtonian gravitational force as $f_{\rm grav}(r)=ME/r^2$
one sees that
$$
{f_{\rm rad}(r)\over f_{\rm grav}(r)}=
{\hbar R^2\over 61,440  \pi^2 M^3 E}   \eqno(ratio)
$$
The size of a macroscopic system always exceeds its Compton length.  Thus
 for any macroscopic sphere  able to fall whole into the
hole $\hbar/E < R < M$.  Therefore, $f_{\rm rad}(r)/g_{\rm rad}(r)\ll 1$
throughout the fall until very close to the hole where the  Newtonian
approximations used must fail.  By then the game is up, and the system must
surely be swallowed up.  It is also clear that the system falls essentially
geodesically (more on this below).

The objection might be refurbished by relying on the radiation pressure of a
large number of  massless species to overpower gravity and
drive the system away.  So let me pretend the number of species in nature is
large.  However, the relevant number, $n$, is the number of species actually
represented in the radiation  flowing out during the time that the
sphere is falling in. I shall take $D$ to be such that the
infall time equals the time $\tau$ to radiate energy $E$.  Then the number of
radiation species into which $E$ is converted is also $n$.    Thus from
\Eq{flux} one sees that the hole radiates the energy $E$ in time $\tau\approx
5\times 10^4 EM^2\hbar^{-1}n^{-1}$.  Since  $D\approx (3\tau/\surd
2)^{2/3}M^{1/3}$, one  checks that $D\approx 2.2\times 10^3
(ME/n\hbar)^{2/3}M$.   Now, the typical Hawking quantum bears an energy of
order $T_{_{BH}}$,
so the number of quanta radiated is  $\approx 8\pi M E/\hbar$.
Since a species will be effective at braking the fall only if  represented by
at least one quantum, one has  $n < 8\pi ME/\hbar$.  As a result,  $D\gg M$ as
required by the discussion.  Multiplying the ratio \(ratio) by $n$ and
recalling that $\hbar/E < R < M$, one sees that
$$
{f_{\rm rad}(r)\over f_{\rm grav}(r)}< {R^2\over 7680  \pi M^2}
\ll 1  \eqno(newratio)
$$
Radiation pressure thus fails to modify appreciably the geodesic fall of
the sphere, and bound \(newbound) follows.

I conclude that the GSL requires for its functioning  a property of ordinary
macroscopic matter encapsulated in bound \(newbound). This is consistent with
the tighter and more definite bound \(mybound) established from statistical
arguments in flat spacetime.  This last granted, the GSL is seen to be safe
from the invasion of a black hole's airspace by  macroscopic entropy--bearing
objects.  It is interesting that this profoundly gravitational law ``knows''
about prosaic physics.  This last statement has been at the heart of a
protracted controversy\refto{UW,BekcUW} in which Unruh and Wald have argued
that the GSL can take care of itself with no help from the entropy bound
by means of the buoyancy of objects in the Unruh acceleration
radiation.  Yet in the {\it gedankenexperiment\/} above  buoyancy is
irrelevant: the sphere falls freely, radiation pressure makes a small
perturbation to its unaccelerated worldline, and so there is no Unruh--Wald
buoyancy.  Evidently, the GSL's functioning does depend on  properties of
ordinary matter.  (For a recent demonstration that the entropy
bound \(mybound) follows from the GSL even in circumstances where
buoyancy is present see \Ref{Bek94} and references cited therein.)

\vglue 0.4cm
\leftline{\twelveit 3.4. Do Black Holes  Emit TV Sets ?}
\vglue 0.3cm

Nothing illustrated so well to my generation the force of the ``no hair''
principle than Wheeler's proverbial TV set falling into a black
hole.\refto{WheelRuff}  But if a black hole can radiate, are TVs emitted in
the Hawking radiance ?  The thermodynamic notion that anything can be found
in a thermal radiation bath would suggest the answer is yes.  This principle,
however, must be applied cautiously.  First, a system of energy $E$ appears
spontaneously in a thermal bath only when the temperature is at least of
order $E$.  A black hole cannot be hotter than the Planck--Wheeler
temperature.  Thus the only TVs that could be expected to appear are
those lighter than the Planck--Wheeler mass $\sim 10^{-5}$ gm.

Further, the TV should be recalcitrant to dissociation.  In the primordial
plasma at redshift $z=10^5$ there were no hydrogen atoms, not because it was
not in equilibrium, but because the corresponding temperature of $3\times
10^5\ {}^0K$ is way above the ioniztion temperature of hydrogen.  There were
$^4$He nuclei then because their dissociation temperature is way above
$3\times 10^5\ {}^0K$.  According to all this logic, Wheeler TVs weighing
much less than the Planck--Wheeler mass, and having a very high dissociation
temperature, should show up in Hawking radiance whose temperature is
of order of the TV's rest energy.  Yet, as I show now,  TV sets or other
macroscopic systems do not occur measurably in any Hawking radiance.

Suppose a macroscopic object (a TV for short)  of size $R$ has rest energy
$E$ and a degeneracy factor $g$.  The last reflects the complexity of the
composite system, so that $g$ could be very large.  The object will get
emitted in an available Hawking mode with probability $g\exp(-E/T_{_{BH}})$.
Actually, if the TV is measurably excited at temperature $T_{_{BH}}$ one should
replace $g$ by an appropriate partition function; I ignore such
complications.   Over the Hawking evaporation lifetime $\sim M^3/\hbar$ there
emerge of order $M^2/\hbar$ ``up'' modes of each species.  Thus the
probability that the hole emits a TV over its lifetime amounts to $p\sim
(M^2/\hbar)g\exp(-8\pi ME/\hbar)$.

Obviously $\ \ln g$ plays the role of internal entropy of the object. From the
bound \(mybound) one may infer that $\ln g < 2\pi RE/\hbar$.  Thus
$p < (M^2/\hbar)\exp[2\pi(R-4 M)E/\hbar]$.  However,  in order for the TV to
be emitted whole it must be smaller than the hole: $R<2M$.  Hence
$p < (M^2/\hbar)\exp(-4\pi ME/\hbar)$.   But obviously the  particles composing
the TV (masses $\ll E$) must have Compton lengths smaller than  $R<2M$ so
that $EM/\hbar\gg 1$.  It follows that the argument of the exponent is very
large, so that $p$ is exponentially small.  Thus in practice an evaporating
black hole does not emit TVs or any macroscopic objects.  This ``selection
rule'' depends on the bound on entropy.

\vglue 0.6cm
\leftline{\twelvebf 4. Acknowledgements}
\vglue 0.4cm

I thank F. Englert, G. Horwitz,  D. Page, R. Parentani, C. Rosenzweig, M.
Schiffer and B. Whiting for critiques and enlightment, U. Bekenstein for
help with the graphics, and the Israel Science Foundation, administered by the
Israel Academy of Sciences and Humanities, for support.

\vglue 0.6cm
\leftline{\twelvebf 5. References}
\vglue 0.4cm

\References

\refis{Bek73}\prd{J. D. Bekenstein}{7}{2333}{73}.

\refis{Carter}\natps{B. Carter}{238}{71}{72}; \inbookeds{B. Carter}{Black
Holes}{B. DeWitt and C. M. DeWitt}{Gordon and Breach, NY}{73}.

\refis{Israel87}\inbookeds{W. Israel}{300 Years of Gravitation}{S. Hawking
and W. Israel}{Cambridge Univ. Press, Cambridge}{87}.

\refis{WaldErice}\inbookeds{R. M. Wald}{Black Hole Physics}{V. de
Sabbata and Z. Zhang}{NATO ASI series, Vol. 364}{92}.

\refis{Israel2}\pla{W. Israel}{A57}{107}{76}.

\refis{Shannon}\book{C. Shannon and W. Weaver}{The Mathematical Theory of
Communication}{University of Illinois Press, Urbana}{49}.

\refis{Srednicki}\prd{M. Srednicki}{71}{66}{93}.

\refis{Bek75}\prd{J. D. Bekenstein}{12}{3077}{75}.

\refis{MullerLousto}\prd{R. M\"uller and C. O. Lousto}{49}{1922}{94}.

\refis{BekPT}\pt{J. D. Bekenstein}{33}{January issue, p. 24}{80}.

\refis{WheelRuff}\pt{R. Ruffini and J. A. Wheeler}{24}{January issue,
p. 30}{71}.

\refis{BCH}\cmp{J. Bardeen, B. Carter and S. W. Hawking}{31}{161}{73}.

\refis{Hawk74}\nat{S. W. Hawking}{248}{30}{74}.

\refis{ParkWald}\prd{L. Parker}{12}{1519}{75}; \cmp{R. M.
Wald}{45}{9}{75}.

\refis{PenrFloyd}\nat{R. Penrose and R. M. Floyd}{229}{177}{71}.

\refis{Hawk75}\cmp{S. Hawking}{43}{199}{75}.

\refis{Chris}\prl{D. Christodoulou}{25}{1596}{70}; \prd{D. Christodoulou and R.
Ruffini}{4}{3552}{71}.

\refis{GibbHawk77}\prd{G. W. Gibbons and S. W. Hawking}{15}{2752}{77}.

\refis{Bek72}\ncl{J. D. Bekenstein}{4}{737}{72}.

\refis{Sorkin86}\prl{R. Sorkin}{56}{1885}{86}.

\refis{Moss}\prl{I. Moss}{69}{1852}{92}.

\refis{Bek74let}\ncl{J. D. Bekenstein}{11}{467}{74}.

\refis{Mukhanov}\jetf{V. Mukhanov}{44}{63}{86}.

\refis{Sch}\preprint{M. Schiffer}{Black Hole Spectroscopy}{ITF, Sao Paulo}{90}.

\refis{Bellido}\preprint{J. Garcia--Bellido}{Quantum Black
Holes}{hep--th/9302127}{93}.

\refis{Maggiore}\preprint{M. Maggiore}{Black Holes as Quantum
Membranes}{hep--th/9401027}{94}.

\refis{Hawklet}\prd{S. W. Hawking}{13}{191}{76}.

\refis{HawkArea}\prl{S. W. Hawking}{26}{1344}{71}.

\refis{York}\prd{J. M. York}{33}{2092}{86}.

\refis{Schiffer93}\prd{M. Schiffer}{48}{1652}{93}.

\refis{Fiola}\preprint{T. M. Fiola, J. Preskill, A. Strominger and S. P.
Trivedi}{Black Hole Thermodynamics and Information Loss in Two Dimensions}
{hep--th/9403137}{94}.

\refis{FrolovPage}\prl{V. P. Frolov and D. N. Page}{71}{3902}{93}.

\refis{FrolovNovikov}\prd{V. Frolov and I. Novikov}{48}{4545}{93}.

\refis{Thorne}\prl{W. H. Zurek and K. S. Thorne}{54}{2171}{85}.

\refis{Jacobson}\preprint{T. Jacobson}{Black Hole Entropy and
Induced Gravity}{gr-qc/9404039}{94}.

\refis{Frolov}\preprint{V. Frolov}{Why the Entropy of a
Black Hole is A/4$}{gr-qc/9406037}{94}.

\refis{Frolov}\preprint{V. Frolov}{Why the Entropy of a Black Hole is
$A/4$ ?}{gr--qc/9406037}{94}.

\refis{BFM}\preprint{A. I. Barvinsky, V. P. Frolov and A. I.
Melnikov}{Wavefunction of a Black Hole and the Dynamical Origin of
Entropy}{gr--qc/9404036}{94}.

\refis{WaldNoether}\prd{R. M. Wald}{48}{3427}{93}.

\refis{PolStro}\preprint{J. Polchinski and A. Strominger}{A Possible
Resolution of the Black Hole Information Puzzle}{Santa Barbara
preprint UCSB-TH-94-20, hep-th/9407008}{94}.

\refis{Visser1}\prd{M. Visser}{46}{2445}{92}.

\refis{Kallosh}\prl{I. Moss}{69}{1852}{92}; \prd{M. Visser}{48}{583}{93};
\prd{R. Kallosh, T. Ortin and A. Peet}{47}{5400}{93}

\refis{JacobsonKang}\prd{T. Jacobson, G. Kang and R. Myers}{49}{6587}{94};
\prd{M. Visser}{48}{5697}{93}.

\refis{Banados}\prl{C. Ba\~nados, C. Teitelboim and J. Zanelli}{72}{957}{94};
\preprint{L. Susskind}{Some Speculations About Black Hole Entropy in String
Theory}{hep--th/9309145}{93}.

\refis{tHooft90}\npb{G. 't Hooft}{335}{138}{90}.

\refis{Sorkin83}\inbookeds{R. D. Sorkin}{General Relativity and
Gravitation}{B.~Bertotti, F. de Felice and A. Pascolini}{Consiglio Nazionale
della Ricerche, Rome}{83}, Vol. 2.

\refis{tHooft93} G. 't Hooft, talk at Santa Barbara Conference
on Quantum Aspects of Black Holes (unpublished, 1993).

\refis{Page93}\prl{D. Page}{71}{3743}{93}.

\refis{Sewell}\pla{G. L. Sewell}{122}{309}{87}.

\refis{Schiffer93}\pr{M. Schiffer}{48}{1652}{93}.

\refis{Dowker}\preprint{J. S. Dowker}{Remarks on Geometric
Entropy}{hep--th/9401159}{94}.

\refis{HolLarWil}\preprint{C. Holzhey, F. Larsen and F. Wilczek}{Geometric and
Renormalized Entropy in Conformal Field Theory}{hep-th/9403108}{94}.

\refis{SusUglum}\preprint{L. Susskind and J. Uglum}{Black Hole Entropy in
Canonical Quantum Gravity and String Theory}{hep--th/9401070}{94}.

\refis{KabStras}\preprint{D. Kabat and D. J. Strassler}{Comment on Entropy
and Area}{hep--th/9401125}{94}.

\refis{CalWil}\preprint{C. Callan and F. Wilczek}{On Geometric
Entropy}{hep--th/9401072}{94}.

\refis{Holzhey} C. Holzhey, Princeton University thesis (unpublished, 1993).

\refis{vonNeumann}\book{J. von Neumann}{ Mathematical Foundations of
Quantum Mechanics}{Princeton Univ. Press, Princeton, N. J.}{55}.

\refis{Wheeler}\book{J. A. Wheeler}{ A Journey into Gravity and
Spacetime}{Freeman, NY}{90}.

\refis{OPenrose}\book{O. Penrose}{ Foundations of Statistical
Mechanics}{Pergamon, New York}{69}.

\refis{STU}\prd{L. Susskind, L. Thorlacius and J. Uglum}{48}{3743}{93}.

\refis{GarfinkleHS}\prd{D. Garfinkle, G. Horowitz and
A. Strominger}{49}{958}{93}.

\refis{Hawk76}\prd{S. W. Hawking}{14}{2460}{76}.

\refis{Hawk82}\cmp{S. W. Hawking}{87}{395}{82}.

\refis{Bek81}\prd{J. D. Bekenstein}{23}{287}{81}.

\refis{StomTriv}\prd{A. Strominger and S. P. Trivedi}{48}{5778}{93}.

\refis{Wil93}\prd{F. Wilczek}{}{}{}.

\refis{Bek74}\prd{J. D. Bekenstein}{9}{3292}{74}.

\refis{Bek93letter}\prl{J. D. Bekenstein}{70}{3680}{93}.

\refis{DanSchiff}\prd{U. H. Danielsson and M. Schiffer}{48}{4779}{93}.

\refis{STU}\prd{L. Susskind, L. Thorlacius and R. Uglum}{48}{3743}{93}.

\refis{Bek84}\prd{J. D. Bekenstein}{30}{1669}{84}.

\refis{BKLS}\prd{L. Bombelli, R. Koul, J. Lee and R. Sorkin}{34}{373}{86}.

\refis{Page80}\prl{D. Page}{44}{301}{80}.

\refis{Page76}\prd{D. N. Page}{14}{3260}{76}.

\refis{Bek79} \inbooked{J. D. Bekenstein}{To Fulfill a
Vision}{Y. Ne'eman}{Addison--Wesley, Reading, Mass.}{81}.

\refis{tHooft85}\npb{G.'t Hooft}{256}{727}{85}.

\refis{BekSchiff90}\ijmpc{J. D. Bekenstein and M. Schiffer}{1}{355}{90}.

\refis{BekSchiff89}\prd{J. D. Bekenstein and M. Schiffer}{39}{1109}{89}.

\refis{BekMeis}\prd{J. D. Bekenstein and A. Meisels}{15}{2775}{77}.

\refis{Giddings92}\prd{S. Giddings}{46}{1347}{92}.

\refis{Bek94}\prd{J. D. Bekenstein}{49}{1912}{94}.

\refis{Giddings94}\prd{S. Giddings}{49}{4078}{94}.

\refis{CGHS}\prd{C.G. Callan, S.B. Giddings, J.A. Harvey, and
A. Strominger}{45}{R1005}{92}; \prd{J. Russo, L. Susskind and L.
Thorlacius}{46}{3444}{92} and {\bf 47} (1993) 533.

\refis{BPS}\npb{T. Banks, M. E. Peskin and L. Susskind}{244}{125}{84}.

\refis{SWJ}\grg{R. Sorkin, R. M. Wald and Z. Z. Jiu}{13}{1127}{81}.

\refis{ZurekPage}\prd{W. H. Zurek and D. N. Page}{29}{628}{84}.

\refis{UW}\prd{W. G. Unruh and R. M. Wald}{25}{942}{82} and
{\bf D27} (1983) 2271.

\refis{BekcUW}\prd{J. D. Bekenstein}{26}{950}{82} and {\bf D27} (1983) 2262.

\refis{Unruh}\prd{W. G. Unruh}{14}{870}{76}.

\refis{PageTmn}\prd{D. N. Page}{25}{1499}{82}.

\endreferences
\endit\end